\documentclass[showpacs,twocolumn,preprintnumbers,amsmath,amssymb,prl,aps]{revtex4}
\usepackage{txfonts}
\usepackage{pifont}
\usepackage{amssymb}
\usepackage{booktabs}
\usepackage{graphicx}%
\usepackage{dcolumn}
\usepackage{amsmath}
\usepackage{epstopdf}
\usepackage{bm}

\hfuzz=\maxdimen
\makeatletter
\def\btt#1{\texttt{\@backslashchar#1}}%
\DeclareRobustCommand\bblash{\btt{\@backslashchar}}%
\makeatother

\begin{document}
	\title{ Forgetting in  order to Remember Better}
	\author{Hang Yu$^\star$}
	\author{Ziyi Liu$^\star$}
	\author{Jiansheng Wu$^\dag$}
	\affiliation{ Shenzhen Institute for Quantum Science and Engineering and Department of Physics, Southern University of Science and Technology, Shenzhen 518055, P.R. China}
	\affiliation{$^\star$These authors contributed equally to this work.}
	\affiliation{$^\dag$ Corresponding author. E-mail: {wujs@sustc.edu.cn}}
	\date{\today}
	
	\begin{abstract}
		In human memory, forgetting occur rapidly after the remembering and the rate of forgetting slowed down as time went. This is so-called the Ebbinghaus forgetting curve. There are many explanations of how this curve occur based on the properties of the brains. In this article, we use a simple mathematical model to explain
the mechanism of forgetting based on rearrangement inequality and get a general formalism for short-term and long-term memory and use it to fit the Ebbinghaus forgetting curve.  
		We also find out that forgetting is not a flaw, instead it is help to improve the efficiency of remembering when human confront different situations by reducing the interference of information and reducing
the number of retrievals. Furthurmove, we find that the interference of information limits the capacity of human memory, which is the ``magic number seven". 
	\end{abstract}

	\maketitle

	In 1885, Herman Ebbinghaus experimentally investigated the properties of human memory quantitatively, and he found that forgetting occur most rapidly after the  remembering and the rate of forgetting slowed down as time went on. He plotted the retention of  nonsense symbols in his memory as a function of time, and this is so-called the Ebbinghaus forgetting
	curve\cite{eb1}.This is the first experiment to investigate the human memory quantitatively. 
	The forgetting curve can be roughly fitted by an exponential function or a power law function quantitatively.   
	On the other hands, modern psychological and neural science show that there are four  mechanism of forgetting: storage failure (the lost of memory mark in neural system), motivated forgetting (forgetting due to emotional reasons, for example, traumatic experiences), interference (failure to recall one information due to the exist of similar informaiton)\cite{inter1,inter2}, retrieval failure (inability to locate a specific memory although it is known to exist).  
	Then an interesting question arise that the exponential decay of human memory are caused by  which of the four possible mechanism, and can we get the forgetting curve
	from these mechanism. The purpose of this article is
	to answer this question.
	
	One simple answer is that the exponential decay is due to the storage failure.
	For example, if the memorized information of a certain situation (We call anythings remembered and to be remembered ``situation", which may include any kinds of events and objects)  $M_{n+1}$ at $n+1$ is proportional to the previous one $M_n$, i.e. $M_{n+1}=s M_n$. And the ratio $s$ is less than one
	due to the lost of memory mark in neural system.  We can get $M_{n+1}=s^n M_{1}=M_1 \exp{\left[- n \ln(1/s) \right]} $. 
	This answer means that
	forgetting is a flaw of neural system. If there are no other advantage of forgetting, such a flaw shouldn't exist
	after handreds of thousand years of evolution of human being. In this article we view the forgetting from another angle: What is the advantage of forgetting?

	Memory is the foundation of human thinking and reasoning. 
	People remember situations  because remembering will increase the efficiency if one has meet these situations
	before then one knows how to  response them in proper ways. In order to  recall corresponding information when confront different situations in
	human lives in a timely and efficient manner, we hope to memorize as much information as possible. But there are two constraints
	on the memory. One is the limitation of  cognitive resource, that is,  human memory has finite capacity. The other is that
	efficiency of retrieval is low and the interference will occur if too many informations are stored. So one best way is to forgetting in order to improve
	the efficiency of retrieval and lower the interference effect. What is the best way to forget  as time going?

	

	\emph{Models.} The basic advantage of memory for human being is that if one remembers certain situation, one know how to response it in a  quick manner, otherwise
	one has to spend more time and effort.  So memory help to improve efficiency.
    But the situation becomes more complicated if one has many informations and one's memory capacity is limited.
	The problem of memory can be modeled as followings:
         How should one use one's finite memory capacity to remember different situations one confronts
         to increase the efficiency to deal with those situations.    
	Here we assume there are $N$ possible situations and the $i$-th situation appear with probability $P_i$ 
	and we have already arrange to probability in descending order ($P_i \ge P_{i+1}$). 
	If one meets a situation, one firstly searches one's memory according the descending order of the possible situations.
	In this way,  one can minimize the expectation of search time. 
	 The time of try-out is,
	\[
	E_{ {T. O.}}=\sum\limits _{i}{i{P_{i}}}\le\sum\limits _{i}{i{P_{{\sigma_{i}}}}}
	\]
	where $P_{\sigma_{i}}$ is a rearrangement of sequence $P_{i} (i=1,2,...,N)$.
	And the inequality is validated due to the rearrangement inequality.\\
	
	
	To illustrate, first we consider
	a binary case with  ${{\rm {P}}_{1}}\ge{{\rm {P}}_{2}}$
	here we have ${E_{\min}}={P_{1}}+2{P_{2}}$.
	Thus if the possibility of ${{\rm {P}}_{1}}\ge{{\rm {P}}_{2}}$ is
	$p$, then the expected value of $E_{\min}$ is,
	\[
	{E_{\min}}=p({P_{1}}+2{P_{2}})+(1-p)(2{P_{1}}+{P_{2}})
	\]
	For cases involving three or more situations, the probability is a bit
	complicated, but we can still extend the above to $N$ possible situation case, 
	\[
	{E_{\min}}=\sum\limits _{all\ \sigma}^ {}{P(}{P_{\sigma_{1}}}\ge {P_{\sigma_{2}}}\ge ...\ge {P_{\sigma_{N}}})\cdot({P_{\sigma_{1}}}+2{P_{\sigma_{2}}}+...+N{P_{\sigma_{N}}})
	\]
	where $\sigma$ represent a cyclic of the elements $1,2,...,N$ and $\sigma_{i}$ represent the $i$-th elements of the cyclic.

	\begin{figure}[tbp]
		\centering
		\setlength{\abovecaptionskip}{2pt}
		\setlength{\belowcaptionskip}{4pt}
		\includegraphics[width=0.8 \columnwidth]{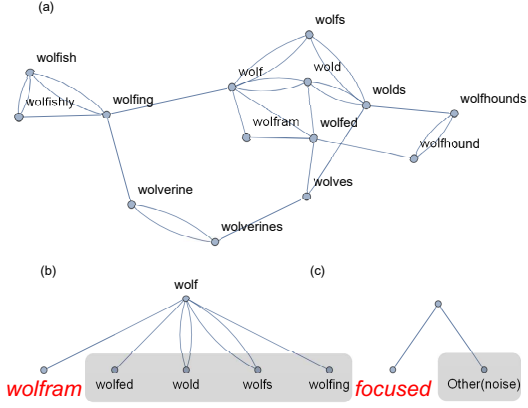}
		\caption{An example of how networks in memory are simplified into a tree and a binary tree.
			In (a) we choose a network of words beginning with \char ```{}wol-\char`\"{},
			in (b) we generate a tree graph that \char```{}wolf\char```{} is chosen
			as root, and  \char```{}wolfram\char`\"{} are chosen as the subject.
			By treating all other situations as noise, we see a binary tree as shown in
			(c).}
		\label{fig.1}
	\end{figure}
	
	In real daily lives of human being,we confront different situations which may form 
	a network instead of a list as above.  But we can use a tree to organize these different situations.
	As shown in Fig(1), for a network  of words beginning with \char```{}wol-\char`\"{},
			we generate a tree graph that  \char```{}wolf\char`\"{} are chosen
			as root and others as the subjects (illustrated in Fig.(1b)). In this way, we have a list of 
			different situations.
          In this article, we mainly consider the binary scenario as it is representative and it is the simplest
	case in mental process. For example, the tree in Fig(1b) can be further simplified as a binary tree:
	The \char```{}wolf\char`\"{}  and \char```{}wolfram\char`\"{}  are chosen as root and subject respectively
	 and all other situations are treated as noise.
	So the  binary tree  is with two branches.
	 one branch (denoted as  situation $f$) is  our focus to which an response can be activated in a proper way,
	and the other branch (denoted as situation $n$) is the noise to which no proper response can be activated.
	
Suppose one meet an incoming situation , and one searches one's memory for situation $1$ (the focused) and then for situation $2$ (the noise) if
	the probability such that $P_f\ge P_n$.  If the situation $1$ is the incoming situation $f$, then he can response in a proper way.
	But if the situation $1$ is not the incoming situation $n$, then he fails to do that. So the probability
	for one to response in a proper way is, 
	\[
	p=\int\limits _{{P_{f}}>{P_{n}}}{\rho({P_{f}},{P_{n}})d{P_{f}}d{P_{n}}}
	\]
	where $\rho({P_{f}},{P_{n}})$ are the joint probability of two situations $f$ and $n$.
	
	Suppose during duration $t$, a situation was recorded $n_{0}^*$ times totally and $n_{0}\ge 1$
	since we need to confront it at lease one time to remember it.
	We call $n_0=n_0^*-1$ the $\it{memory\ frequency}$ and it is one of the elementary factors which affect the strength of memory positively. 
	In most of the case in daily lives,  ${n_{0}}$ satisfies the Poisson distribution with frequency
	$\lambda$, i.e.,  
	\[
	P({n_{0}}|{ \lambda})=\frac{{{{(\lambda t)}^{{n_{0}}}}{e^{-\lambda t}}}}{{\Gamma({n_{0}}+1)}}.
	\]
	
	But the frequency $\lambda$ is unknown. Using the Bayes statistics with a constant prior distribution $\rho(\lambda )$, we can
	estimate $\lambda$ from $n_0$ by
	\[
	P\left({{\lambda}|{n_{0}}}\right)=p({n_{0}}|{\lambda})\rho({\lambda})/{\rho({n_{0}})}=\frac{{{\lambda^{{n_{0}}}}{t^{{n_{0}}+1}}{e^{-\lambda t}}}}{{\Gamma({n_{0}}+1)}}
	\]
	
	Here, in the binary tree, the branch of the focus is the incoming situation with frequency $\lambda$ and the branch of noise is a noise assumed to appear at frequency $k$, so the probability that the situation can be recalled, is  the followings,
	\begin{eqnarray}
	\begin{array}{l}
	P_{\rm M}(n_0, kt)=\int_{k}^{\infty}{P({\lambda}|{n_{0}})d{\lambda}}={\Gamma({n_{0}}+1,kt)}/{\Gamma({n_{0}}+1)}\label{pM}
	\end{array}
	\end{eqnarray}
	where $\Gamma({n_{0}}+1,kt)=\int_{k}^{\infty} {{\lambda^{{n_{0}}}}{t^{{n_{0}}+1}}{e^{-\lambda t}}} d \lambda=\int_{kt}^{\infty} x^{(n_0+1)-1} e^{-x} dx $ 
	is the upper incomplete $\Gamma$-function and the $P_M$ function is the regularized 
	upper incomplete $\Gamma$-function. This function is actually the probability over a duration time $t$ of which
	a situation (which has appeared $n_0$ times) can be recalled.  $P_{M}$ is equal to $1$ at the beginning ($t=0$),
	It work as the retention strength of memory as a function of time, i. e., the memory function. 
	And it equals to $1$ if there exist no noise ($k=0$) as well. We can see here the forgetting of memory
	partially are due to the interference of noise. 
	
	If an event is only encountered at the initial and doesn't appear again, i.e. $n_{0}=0$, 
	then we have $p_M=e^{-kt}$ which is the exponential function usually used to fit the Ebbinghaus forgetting curve. Here $k$ is the frequency of the noise event (assumed to satisfy the Poisson process as well). And we can
	see the larger the frequency of the noise events, the faster the memory strength decay. We can see from Eq.($\ref{pM}$) that the larger the $n_0$, the slower the decay ratio at small $t$
	which means that the memory strength can last for a longer time. For much larger $n_0$,
	we can see the memory last longer which is still around $1$ when $kt \le n_0$. $P_{M}(k, t)$ can be expanded as,
	\[
	P_M(n_0,kt)=\frac{\Gamma(n_0+1, kt)}{\Gamma({n_0+1}) } =  e^{-kt}\sum_{s=0}^{n_0}\frac{(kt)^s}{\Gamma(s+1)}.
	\]
	which is the cumulative distribution function for Poisson random variables: If $X$ is a ${\rm Poisson} (k)$ random variable with frequency $k$ then
	$ P(X< n_0+1) =  P_M(n_0,kt )$. So we can interpretate the forgetting curve in another way:
	If  a situation applear $n_0+1$ times and $n_0+1$ is greater than the number of time that noise situation appear, we can remember it.
	Furthermore, from the above reasoning,  we can see how the memory function behavior depend on the distribution
	of the incoming situations. For different distribution, we can get different memory function. 
	
	If the incoming situation appear according the binomial process with total number of
	sampling $N=t/\tau$ and the situations occur $n_0+1$ times with probability $k\tau$ ($\tau$ is the duration when a situation appear and
	$k$ is the frequency of the noise situation), then the memory function should be
	\[
	P^{B}_M(n_0,k,\tau, t)=\frac{B( k\tau, n_0+1,N-n_0-1)}{B(n_0+1,N-n_0-1) } .
	\]
	where $B(a,b)$ is the Beta function and $B(x,a,b)$ is the incomplete Beta function.

Some situations are special. For examples, some situations happen one or several times in a life time (for examples, marriage), 
	they appear with probability $p$ and it doesn't varies with
	time. 	 The forgetting curve is
	\[
	P^{C}_M(p, t)=\Theta( p-k).
	\]
	where $k$ is probability of  such kind of situations other than the focused one and $\Theta(x)$ is a step function s. t. $\Theta(x)=1$ for $x\ge 0$ and $\Theta(x)=0$ for other case. 
	So the memory function is a constant. Once such situation happen, it is remembered and it is forgotten once a new noise situation of such kind happen. 
	

	It was believed that long-term memory is from the consolidation of short-term memory, so-called memory consolidation\cite{Consolidation2}.
	However, different from Atkinson Shiffrin memory model\cite{spe2}, studies of patients with
	perisylvian cortex damage and inferior parietal cortex  damage show that these patients had deficits of short-term memory 
	but long-term memory was preserved\cite{spe3,sp1,sp2}. If  memory consolidation of short-term memory is the only way to form long-term memory, 
	the above situation shouldnot   exist.
	And the protein kinase C, zeta (PKC zeta)\cite{protein} which might be important for creating and maintaining long-term memory 
	and not important for short-term memory suggests that long-term memory is different from short-term memory.  
	So there are three parts of memory, including short-term memory (it is also called working memory),
	 long-term memory not from memory consolidation and long-term memory from consolidation (memory consolidation)\cite{consolidation}.
	Here we can use $P_M$ with small $n_0$ as short-term memory
	function and those with large $n_0$ as long-term memory function. At the same time, they can have different
	value of noise frequency $k$. Then it is reasonable that we assume arbitrary memory function (or forgetting curve) can be decomposed as
	\begin{equation}
	P_{LM}(k,t)=\sum_{n=1}^\infty C_{n}\frac{\Gamma(n+1,kt)}{\Gamma(n+1)}
	\end{equation}
	$C_n$  are constant s.t.  $\sum_{n=1}^\infty C_{n}$=1 
	 which grantee the unitary at $t=0$.
	To fit Ebbinghaus forgetting curve, we need two sets of $n_0$ and $k$ values and  three forgetting functions to well fit the curve shown in the next section.

	
	For long-term memory, we take relative small constant $ k_{l}$ because the noise of environment is stabilized at long time scale. Based on the study of interference theory\cite{interfe,interfe2,inter2} and serial-position effect\cite{serial}, the former and latter information can affect the memory of middle information negatively. 
	We consider there is an increment of noise $k_{s}$ for short-term memory and when more situations are confronted.
	Assuming that the noise frequency is $k_{s0}$  for short-term memory for first incoming situation,
	the environmental noise for the $i$-th incident situation is $i k_{s0}$.
  	
	Using the memory function with $n_0=0$ as an approximation, the expected number of situation (memory capacity) that can be remembered  during  $\Delta$  and the total number of try-out are the followings respectively, 
	\[
	{E_{C}}=\sum\limits _{i=0}^{\infty} p_{i}=\sum\limits _{i=0}^{\infty}{e^{-ik_{s0}\Delta}}=\frac{1}{(1-{e^{-k_{s0}\Delta}})}
	\]
	\[
	{E_{T.O.}}=\frac{\sum \limits _{i=0}^{\infty}{ip_{i}}}{\sum\limits _{i=0}^{\infty}{ p_{i}}}=\frac{{\sum\limits _{i=0}^{\infty}{i{e^{-ik_{s0}\Delta}}}}}{{\sum\limits _{i=0}^{\infty}{e^{-ik_{s0}\Delta}}}}=\frac{{e^{-k_{s0}\Delta}}}{{(1-{e^{-k_{s0}\Delta}})}}=(E_C-1)
	\]
	It is interesting to notice that the try-out numbers
	is increasing with the memory capacity expanding.
At the same time to reduce the number of try-out, the working
	space should be limited so that one can response to a situation in a quick 
	and efficiency manner.

	
	\emph{Results.} The first experimental study on human memory was done
	in late 19th century by Ebbinghaus.  It demonstrated the basic characteristics
	of how memories fading over time. As is proposed in the article, 
    Ebbinghaus chose exponential
	model to fit his results, a power law and later logarithm law.
And better fitting models are appearing since then\cite{eb1,Eb2}.
	Although the experiment used percentage of time reduction instead
	of percentage of retention, here we directly take the two representation
	of memory strength as equal. 
	Here we use our model to fit the data of experiments by Ebbinghaus\cite{eb1}, Mack and Seitz\cite{seitz} and Dros\cite{Eb2} in Fig.3.
	Our function is 
	\[
	p=0.68\frac{\Gamma(n_{s}+1,k_{w0}t)}{\Gamma(n_{s}+1)}+0.13\frac{\Gamma(n_{s}+1,k_{l}t)}{\Gamma(n_{s}+1)}+0.19\frac{\Gamma(n_{l}+1,k_{l}t)}{\Gamma(n_{l}+1)}
	\]\\
	Here we choose two $n_0$ and two $k$ values to construct the function of forgetting curve.
	That is, $n_{s}=0$ characterizes working memory, $n_{l}=13$ characterizes the long-term memory, $k_{s0}=300/day$, and $k_{l}=1/day$. 
	We can see long-term memory is characterized by larger $n_0$ and lower noise frequency
	and the short-term memory is opposite.
	The second term is for the memory consolidation, which means the transformation from short-term memory to long-term memory as
	mentioned above. It is very interesting that in this memory consolidation term,  a good fitting need a low noise frequency  as long-term
	memory and  a $n_0$ value as short-term memory instead of a noise frequency as
	short-term memory and a $n_0$ value as long-term memory.  It is counterintuitive. The intuitive thinking  is that 
	the repeating of short-term memory is to  increase the $n_0$ value.  But it is not the case here.
	The fucntion of  the repeating of the short-term memory is to
	lower the noise frequency from short-term memory level to long-term memory level.
	\begin{figure}[tbp]
		\centering
		\setlength{\abovecaptionskip}{2pt}
		\setlength{\belowcaptionskip}{4pt}
		\includegraphics[angle=0, width=1.4 \columnwidth]{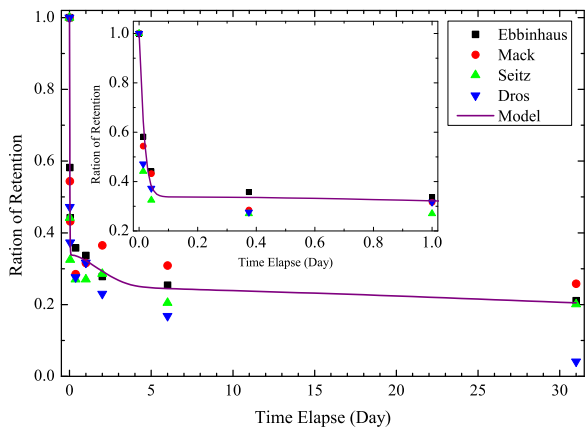}
		\caption{ The fitting of 
the experimental data of Ebbinghaus, Mack and Seitz and Dros using our model. 
The inserted small picture is details between 0 and 1.}
		\label{fig:PhaseDiag-E-B}
	\end{figure}
	
	For memory capacity of short-term memory, 
we take $k_{s0}=300/day$ from function $p(t)$ and the increment of $k_{s}$ happens  during a short time scale in our model. The  period of short-term memory $\Delta$ can be considered as the duration of the attention or test\cite{mi}, so we take $\Delta$ from $30$ seconds to $1$ min.
	We get ${E_{C}}\in[4.3, 9.1]$. This is close to ``the magic number $7$"  which is the average number of non-correlated items that people can remember in short-term memory\cite{Miller}.
	
	In conclusion,  by using a simple rearrangement inequality, 
	we get a general formalism for both short-term and long-term memory function.
	How fast the forgetting occur depends on the number that a remembered situation occurs and the
	noise frequency of the noise situation.  The Ebbinghaus's forgetting curve can be well fitted
	by three memory function: short-term memory, long-term memory and memory consolidation.
	The memory consolidation term shows that the impact of repeating of short-term memory
	is to reduce the noise frequency. 
	The forgetting is not a flaw, on the contrary, it can help to increase the efficiency of remembering
and this is consistent with some recent psychological experiments\cite{forgetting}. It has two folds of impacts.
	One is that it reduce the interference of  informations. 
	The second is that it reduces the number of retrievals. It is shown in our paper that the interference
	of information limits the capacity of memory and  such number is around seven, so-called the
	magic number seven. The number of try-out to recall a certain information is
	proportional to the capacity of the memory, so by lower the capacity, we can recall
	information more quickly.

\end{document}